\documentclass[conference,hidelinks]{IEEEtran}
\IEEEoverridecommandlockouts
\usepackage{cite}
\usepackage{amsmath,amssymb,amsfonts}
\usepackage{algorithmic}
\usepackage{graphicx}
\usepackage{textcomp}
\usepackage{xcolor}
\usepackage{tablefootnote}
\usepackage{placeins}

\ifCLASSOPTIONcompsoc
    \usepackage[caption=false, font=normalsize, labelfont=sf, textfont=sf]{subfig}
\else
\usepackage[caption=false, font=footnotesize]{subfig}
\def\BibTeX{{\rm B\kern-.05em{\sc i\kern-.025em b}\kern-.08em
    T\kern-.1667em\lower.7ex\hbox{E}\kern-.125emX}}
\makeatletter
\newcommand{\linebreakand}{%
  \end{@IEEEauthorhalign}
  \hfill\mbox{}\par
  \mbox{}\hfill\begin{@IEEEauthorhalign}
}
\makeatother

\begin{document}

\title{SOS: A Shuffle Order Strategy for Data Augmentation in Industrial Human Activity Recognition
}

\author{\IEEEauthorblockN{1\textsuperscript{st} Anh Tuan Ha\IEEEauthorrefmark{1}}
\IEEEcompsocitemizethanks{\IEEEcompsocthanksitem\IEEEauthorrefmark{1}Equal contribution}
\IEEEauthorblockA{\textit{Faculty of Computer Science} \\
\textit{and Engineering}\\
\textit{Ho Chi Minh city}\\
\textit{ University of Technology-}\\
\textit{Vietnam National University}\\
Ho Chi Minh City, Vietnam  \\
tuan.haanhego@hcmut.edu.vn}
\and
\IEEEauthorblockN{2\textsuperscript{nd} Hoang Khang Phan\IEEEauthorrefmark{1}}
\IEEEauthorblockA{\textit{Department of Biomedical} \\
\textit{Engineering}\\
\textit{Ho Chi Minh city}\\
\textit{ University of Technology-}\\
\textit{Vietnam National University}\\
Ho Chi Minh City, Vietnam\\
khang.phan2411@hcmut.edu.vn}
\and
\IEEEauthorblockN{3\textsuperscript{rd} Thai Minh Tien Ngo }
\IEEEauthorblockA{\textit{Faculty of Computer Science} \\
\textit{and Engineering}\\
\textit{Ho Chi Minh city}\\
\textit{ University of Technology-}\\
\textit{Vietnam National University}\\
Ho Chi Minh City, Vietnam \\
tien.ngozack2004@hcmut.edu.vn}
\linebreakand
\IEEEauthorblockN{4\textsuperscript{th} Anh Phan Truong }
\IEEEauthorblockA{\textit{Faculty of Computer Science} \\
\textit{and Engineering}\\
\textit{Ho Chi Minh city}\\
\textit{ University of Technology-}\\
\textit{Vietnam National University}\\
Ho Chi Minh City, Vietnam\\
phan.truongbraddock@hcmut.edu.vn}

\and
\IEEEauthorblockN{5\textsuperscript{th} Nhat Tan Le }
\IEEEauthorblockA{\textit{Department of Biomedical} \\
\textit{Engineering}\\
\textit{Ho Chi Minh city}\\
\textit{ University of Technology-}\\
\textit{Vietnam National University}\\
Ho Chi Minh City, Vietnam\\
lenhattan@hcmut.edu.vn}
}
\IEEEoverridecommandlockouts
\IEEEpubid{\makebox[\columnwidth]{979-8-3503-7550-3/24/\$31.00 \copyright2025
IEEE \hfill}
\hspace{\columnsep}\makebox[\columnwidth]{ }}
\maketitle
\IEEEpubidadjcol
\begin{abstract}
In the realm of Human Activity Recognition (HAR), obtaining high quality and variance data is still a persistent challenge due to high costs and the inherent variability of real-world activities. This study introduces a generation dataset by deep learning approaches (Attention Autoencoder and conditional Generative Adversarial Networks). Another problem that data heterogeneity is a critical challenge, one of the solutions is to shuffle the data to homogenize the distribution. Experimental results demonstrate that the random sequence strategy significantly improves classification performance, achieving an accuracy of up to 0.70 \textpm{0.03} and a macro F1 score of 0.64 \textpm{0.01}. For that, disrupting temporal dependencies through random sequence reordering compels the model to focus on instantaneous recognition, thereby improving robustness against activity transitions. This approach not only broadens the effective training dataset but also offers promising avenues for enhancing HAR systems in complex, real-world scenarios.
\end{abstract}


\section{Introduction}
\label{section:Introduction}
Human Activity Recognition (HAR) is a critical area of research with significant applications in industrial automation, healthcare monitoring, and smart environments. In manufacturing and logistics settings, accurately recognizing human activities can lead to improved efficiency, better workflow optimization, and enhanced safety measures. However, a major challenge in HAR is the collection of high-quality, diverse datasets that truly capture the variability of human actions in real-world scenarios. The costs associated with large-scale data collection, the complexity of human movements, and the difficulty in labeling time-series data limit the effectiveness of traditional HAR models\cite{tee2022closelookhumanactivity}.
In logistics centers, workers engage in sequential and repetitive tasks, such as scanning labels, assembling boxes, and packaging items. These tasks often vary depending on product size, worker technique, and workflow disruptions, making it difficult for standard HAR models to generalize effectively.

In addition, the ability to accurately recognize these activities is crucial. It empowers employers to optimize workflow management, enabling real-time adjustments for efficiency and resource allocation. Furthermore, it facilitates rapid error detection, minimizing costly mistakes like mislabeling or incorrect packaging.

Nevertheless, the scarcity of labeled data restricts the development of robust classification models, which struggle to adapt to unseen variations in human activities\cite{info10080245}. Existing data augmentation strategies, such as synthetic data generation, provide some relief, but they often fail to fully capture real-world complexities or improve model generalization.

To address this problem, this study introduces a novel strategy for HAR classification applied specifically to the OpenPack dataset\cite{yoshimura_2023_8145223} for ABC Challenge 2025- Virtual Data Generation
for Complex Industrial Activity Recognition. Utilizing sensor and operations data, our approach integrates some deep learning-based synthetic data generation combined with a strategic data augmentation method to enhance model performance and generalization.

\section{Related Works}
\label{section: Related literature}
The field of Human Activity Recognition (HAR) has significantly advanced with the integration of sensor-based and deep learning techniques. Early works in HAR primarily utilized traditional machine learning algorithms such as Support Vector Machines (SVM) or Extreme Gradient Boosting Classifier to classify human activities based on wearable sensor data \cite{10.1145/3675094.3678461, 8231736, 8537550, s21186316}. However, these methods relied heavily on handcrafted features, limiting their adaptability to real-world environments with varying conditions and sensor noise. To overcome this, deep learning techniques such as Recurrent Neural Networks (RNNs) and Long Short-Term Memory (LSTMs) were introduced, offering improved accuracy by capturing temporal dependencies in sequential activity data \cite{8843403}. More recently, Transformer-based HAR models have emerged, leveraging self-attention mechanisms to enhance feature extraction and activity classification \cite{leite2024transformerbasedapproachessensorbasedhuman}.

Parallel to these advancements, the challenge of data scarcity in HAR has driven the exploration of data augmentation and synthetic data 
generation techniques. Generative models such as Conditional Generative Adversarial Networks (CTGANs) and Variational Autoencoders (VAEs) have been widely used to generate realistic sensor data, improving model generalization when training on small or imbalanced datasets \cite{ 8285168}, which improve the model classification to 0.9386 F1 score. Studies by Parisa Fard Moshiri et al. \cite{moshiri2020usingganenhanceaccuracy} have demonstrated that synthetic data augmentation can enhance classification accuracy in HAR tasks by introducing greater variability in the training data, this results in a 3.4\% increase in classification accuracy and a 15\% reduction in log loss. Despite these improvements, many existing methods fail to capture the sequential nature of real-world human activities fully, often leading to inconsistent synthetic data distributions that hinder classification performance.

Building on this foundation, our study introduces a novel data augmentation strategy that integrates deep learning-based synthetic data generation with strategic sequence reordering. Unlike traditional augmentation methods that either preserve strict sequence order or randomly shuffle data without contextual guidance, our approach leverages Attention Autoencoder (AAE) and CTGAN models to generate realistic sensor data, which is then strategically reordered using a Shuffle Data Augmentation Strategy. This approach aims to (1) mitigate the limitations of conventional augmentation techniques, (2) enhance model generalization in HAR classification, and (3) benchmark against existing augmentation methods to evaluate its effectiveness in improving activity recognition performance.

\section{Methodology}
\label{section:Methodology}
In this study, we utilize the OpenPack dataset\cite{yoshimura_2023_8145223, xia2024preliminaryinvestigationsslcomplex, 10214902}, which comprises 21 workers packaging delivery boxes. Firstly, to augment the generalizability of training synthesizer data, we select two random people from the dataset and make a combined dataset comprised of the worker accelerometer data from those 2 people. Then, for sequence setting, we employ a random label order base for a time series data generation strategy, which will be compared with two baseline methods - activity ascending ordering and reshaping the generated data 
(see subsection \ref{Synthetic data strategy}). The strategy was tested in three augmentation conditions (combined dataset, CTGAN-generated dataset \cite{xu2019modelingtabulardatausing}, and AAE-generated dataset)


\begin{figure}[!ht]
    \centering
    \includegraphics[width=\linewidth]{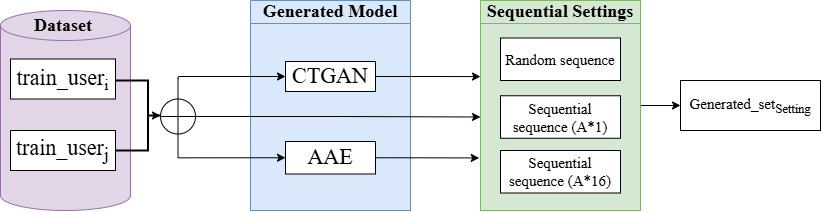}
    \caption{The proposed pipeline, where i and j are two random persons from the training dataset; A represents the activity sequence from activity 100 to 1000.}
    \label{fig:challenge3}
\end{figure}
\FloatBarrier

\subsection{Dataset}
\label{Dataset}
OpenPack \cite{yoshimura_2023_8145223, xia2024preliminaryinvestigationsslcomplex, 10214902} is a large-scale, multimodal dataset designed to comprehensively capture the complexity of real packaging operations in modern logistics centers. This dataset was collected in a dedicated simulated environment constructed within a 3m × 5m space that closely replicates the workspace found in warehouses. The dataset comprises a total of 53.8 hours of sensor data recorded from 104 collection sessions, with 20,161 labeled instances of operations and 53,286 labeled instances of actions. The packaging operations are categorized into 10 main activity classes, while the finer-grained actions are divided into 32 classes, detailing each step in the order processing workflow—from product selection and inspection to box assembly, labeling, and finishing in the table \ref{tab:op-steps-4x6}.
\begin{table}[!ht]
\caption{The table of operation activities.}
\centering
\begin{tabular}{clcl}
\hline
ID & Operation & ID & Operation \\
\hline
100 & Picking                &700 & Scan Label  \\
200 & Relocate Item Label    & 800 & Attach Shipping Label \\

300 & Assemble Box           &900 & Put on Back Table  \\

400 & Insert Items           & 8100 & Others  \\

500 & Close Box              & 1000 & Fill out Order \\

 600 & Attach Box Label  &                        &  \\
\hline
\end{tabular}

\label{tab:op-steps-4x6}
\end{table}

\begin{figure}[!ht]
    \centering
    \includegraphics[width=\linewidth]{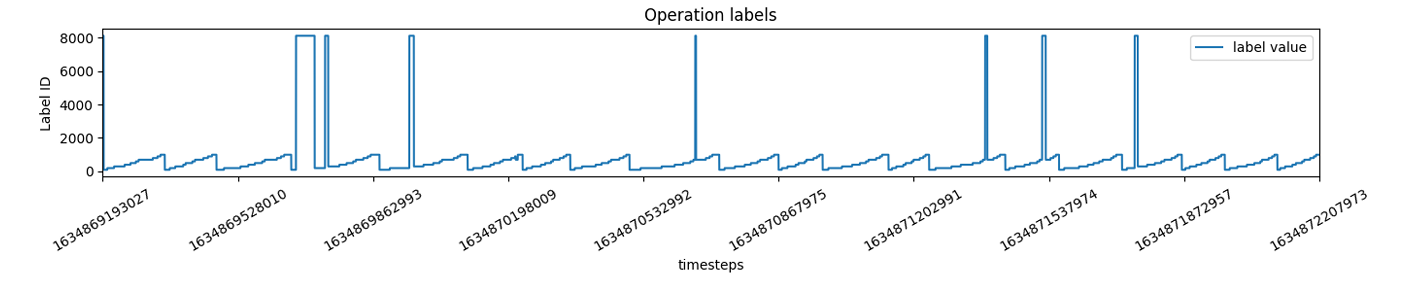}
    \caption{The figure of operation distribution}
    \label{fig:rawdata}
\end{figure}

As in Figure 2, we have a clear hierarchical labeling structure for 10 primary operations. These labels span the entire packaging workflow, from initial steps (e.g., picking, relocating, etc.) to final stages (e.g., Put on back table, fill out, ...). This multi-level classification not only highlights distinctions among the various phases of the process but also supports more fine-grained analyses.
In addition, the figure \ref{fig:rawdata} depicts the raw sensor data associated with these operations, illustrating how signal variations over time reflect the actual state of the packaging process.

Upon analyzing the dataset, we observed that label 8100 constitutes less than 5\% of the overall instances. Given its limited representation, we determined that further processing or targeted augmentation for this particular class was unnecessary. Preserving its original state helps to maintain the natural distribution of the dataset and avoids potential overfitting or the introduction of bias that might arise from artificially inflating this underrepresented category.\\

\subsection{Activity recognition model} 
\label{Validation model}
\begin{figure}[!ht]
    \centering
    \includegraphics[width=\linewidth]{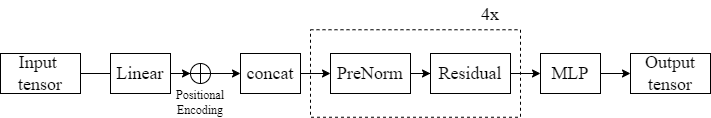}
    \caption{The architecture of discriminative model}
    \label{fig:HAR_model}
\end{figure}
Human Activity Recognition (HAR) has become increasingly vital with the rapid expansion of wearable technologies and IoT devices. A Transformer-based approach to HAR that leverages high-dimensional sensor data to accurately classify human activities.  

In figure \ref{fig:HAR_model}, the process begins with data pre-processing: The model begins to collect data from OpenPack datasets. Each data file includes multiple sensor readings, typically the x, y, and z acceleration values, along with the operational labels. Once imported, these datasets are converted into NumPy arrays. In addition, activity labels are standardized by encoding them in numerical values, ensuring that all parts of the data set share a consistent format. 

Next, to capture the temporal dynamics inherent in time-series sensor data, the model employs a sliding window technique. The continuous sensor data is segmented into fixed-length sequences using a window size of 300 time steps, with each segment overlapping by 150 steps. This overlapping strategy ensures that transitional information between different activities is preserved. 
This encoding, implemented through sine and cosine transformations, embeds the time-step position into the data, thus maintaining the order of events and enabling the model to learn temporal dependencies more effectively. 

At its core, the HAR model leverages a Transformer\cite{vaswani2023attentionneed} architecture. The process starts by projecting the segmented sensor data into a higher-dimensional embedding space using a linear layer. A classification token is then appended to the sequence to aggregate global context. These blocks work in tandem to capture intricate patterns and temporal relationships within the data. The final output from the Transformer is fed into a classifier that generates the predicted activity labels. The training phase uses a cross-entropy loss function optimized by the Adam optimizer with a Cosine Annealing learning rate scheduler. Furthermore, early stopping is integrated to stop training when the validation loss ceases to improve, thus preventing overfitting. Finally, the performance of the model is rigorously evaluated in a separate test set using the macro-averaged precision, recall, F1 score, and accuracy.

\subsection{Synthetic data models}
\label{Synthetic data model}
CTGAN\cite{xu2019modelingtabulardatausing}, short for the Conditional Tabular Generative Adversarial Network, is an advanced methodology crafted to tackle the inherent challenges of modeling and synthesizing tabular data. Unlike conventional approaches, CTGAN is deliberately designed to manage the complexities associated with heterogeneous data types, including both continuous and discrete variables. 
Although it was originally designed for tabular data, CTGAN’s advanced architecture makes it equally effective for generating high-quality time series data. Its ability to handle both continuous and discrete variables ensures that intricate data patterns are accurately captured, while its conditional generation feature empowers users to steer the synthesis process precisely. This means that critical temporal dependencies and variability in time series are faithfully reproduced, enabling more robust and reliable predictive models. In essence, CTGAN offers a persuasive solution for creating realistic synthetic data where meticulous analysis and data quality are paramount. In this paper, we train CTGAN from SDV \cite{SDV} with 3 epochs for this CTGAN data generation.
 
\label{Validation model}
\begin{figure}[!ht]
    \centering
    \includegraphics[width=\linewidth]{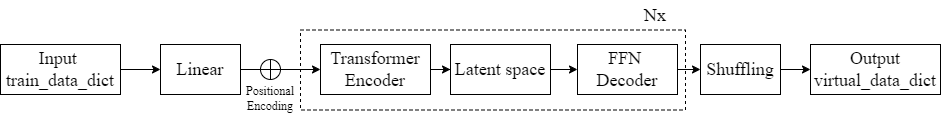}
    \caption{The architecture of proposed AAE}
    \label{fig:AAE}
\end{figure}
AAE integrates the attention mechanism with an autoencoder (AE)\cite{michelucci2022introductionautoencoders} to optimize the extraction of information from complex data. The model constructs a probabilistic latent space that ensures both smoothness and precision in data reconstruction while also enabling the generation of synthetic samples with controlled randomness. The attention mechanism allows AAE to identify critical contextual relationships by dynamically allocating weights among the elements of a data sequence, thereby enhancing the encoding process and reducing information loss. This integration of the encoding-decoding phases with attention not only improves the handling of high-dimensional, large-scale datasets but also broadens the model’s applicability to fields such as natural language processing, synthetic image generation, and time series analysis. Figure \ref{fig:AAE} illustrates the architecture of AAE, clearly demonstrating how the key components are interconnected in the process of information learning and synthesis. Specifically, with AAE, we trained the model with a learning rate of 0.001 and 100 epochs to generate data.
\subsection{Synthetic data strategy}
\label{Synthetic data strategy}
In data generation, models typically aim to capture and resample the patterns present in the training data. A diverse and representative training dataset is crucial for maximizing the variability of the synthetic data generated by the model. To enrich the training data and promote greater diversity in the generated synthetic data, we combined data from multiple individuals. This approach allowed the model to learn a wider range of patterns and variations in human behavior.

Then, the combined dataset was used to train a generated model for synthetic data generation. The resulting synthetic data was then reshaped according to the three settings described below for evaluation:

\begin{itemize}
    \item Random sequence (RS):  In the random sequence setting, the order of the sequences in the dataset was shuffled. This disrupted the temporal dependencies in the data, forcing the classification model to make predictions based on individual time steps rather than sequence context.
    \item Ascending sequence (AS): In the sequential sequence setting, the dataset was organized into sequences of consecutive, ascending activity labels. This minimized transitions between activities, allowing the model to learn activity patterns with minimal noise from transitions.
    \item Real dataset sequence (RDSS): The real dataset sequence setting aimed to mimic realistic activity patterns. The dataset was first shuffled and then divided into 16 groups, which were then arranged sequentially. Within each group, sequences were created using the same ascending activity label strategy as in the ascending setting. These sequences were then concatenated to form a dataset with realistic activity transitions.
\end{itemize}

As a baseline, the untouched combined dataset was also reshaped according to the ascending sequence and random sequence settings described above. Additionally, we also train a model without data augmentation (WDA) for testing the overall impact of generated data to the recognition task. This allowed us to compare the performance of models trained on the original data with those trained on the synthetically generated data, thus evaluating the impact of the synthetic data generation process.

\section{Results and Analysis}
\label{section:Result}

\begin{table*}[!ht]
\caption{The classification result of all settings and baseline (in mean (std))}
    \centering
    \begin{tabular}{p{0.1\textwidth}
p{0.055\linewidth}p{0.055\linewidth}p{0.055\linewidth}
p{0.055\linewidth}p{0.055\linewidth}p{0.055\linewidth}
p{0.055\linewidth}p{0.055\linewidth}p{0.055\linewidth}
}
        \hline
    Model&\multicolumn{3}{c}{{\textbf{CTGAN}}}&\multicolumn{3}{c}{{\textbf{AAE}}}&\multicolumn{3}{c}{{\textbf{Original Data}}}\\

      Setting   & RS & AS & RDSS & RS & AS & RDSS  & RS & AS & WDA \\
    \hline
       Accuracy  & \textbf{0.70 (0.03)} & 0.63 (0.01) & 0.62 (0.05) &0.67 (0.06)  &0.61 (0.04)  &  0.60 (0.05) & 0.65 (0.00) & 0.63 (0.06) &  0.68 (0.01)\\
       \hline
     Precision    & 0.64 (0.01) & 0.59 (0.01) & 0.59 (0.03)  &\textbf{0.69 (0.08)}  &0.57 (0.04)  &  0.57 (0.04) &0.61 (0.02)  & 0.59 (0.05) &  0.64 (0.02)\\
     \hline
      Recall   &0.61 (0.01)  & 0.53 (0.00) & 0.54 (0.04)  &\textbf{0.62 (0.08)}& 0.56 (0.04)  & 0.55 (0.06) & 0.60 (0.01) & 0.56 (0.05)  &  0.60 (0.00)\\
      \hline
      \textbf{Macro F1}   & 0.63 (0.02) &0.54 (0.01)  &0.54 (0.05)  &\textbf{0.64 (0.01)}  &0.56 (0.04)  & 0.54 (0.06) & 0.60 (0.01) & 0.56 (0.05) &  0.61 (0.00)\\
      \hline
    \end{tabular}
    
    \label{tab:my_table}
\end{table*}

The experimental results are summarized in Table \ref{tab:my_table}, which reports the classification performance of three approaches: CTGAN, AAE, and Untouched DF (original data).

In the study, the AAE model with the RS setup demonstrates remarkable performance characteristics. Specifically, although the accuracy reached 0.67 with a high standard deviation (0.60)—indicating significant variability in prediction capability—the precision, recall, and Macro F1 scores are impressive, with values of 0.69 \textpm0.08, 0.62 \textpm 0.08, and 0.64 \textpm 0.01, respectively. This reflects the model’s ability to correctly identify classes and enhance overall classification performance. These results suggest that, although additional measures are needed to mitigate the variability in accuracy, the AAE RS model still holds considerable promise for future classification applications.
The CTGAN model under the RS configuration exhibits the highest mean accuracy at 0.70 \textpm 0.03, suggesting that the synthetic data generated by CTGAN can lead to robust classification performance. However, its precision 0.64 \textpm 0.01, recall 0.61 \textpm 0.01, and Macro F1 score 0.63 \textpm 0.02 are slightly lower than those observed for the AAE model in the RS setting. This indicates that while CTGAN provides stability in overall accuracy, the ability to correctly classify individual classes might benefit from further refinement.

The Untouched DF approach, using the original data with the WDA configuration, also demonstrates good performance. With an accuracy of 0.68 \textpm0.01, a precision of 0.64 \textpm0.02, and a Macro F1 score of 0.61 \textpm0.00, it confirms that traditional methods, when combined with effective weighting strategies, can serve as a reliable benchmark for synthetic data approaches. Although both approaches yield comparable mean accuracy, the high standard deviation in the AAE RS setup of 0.06 contrasts sharply with the stable performance of the Untouched DF method of 0.01. This suggests that while AAE RS may achieve higher precision (0.69 vs. 0.64) and slightly better recall and Macro F1 scores, its reliability is hindered by considerable variability in overall accuracy.

\begin{figure*}[!ht]
    \centering
    \includegraphics[width=\linewidth]{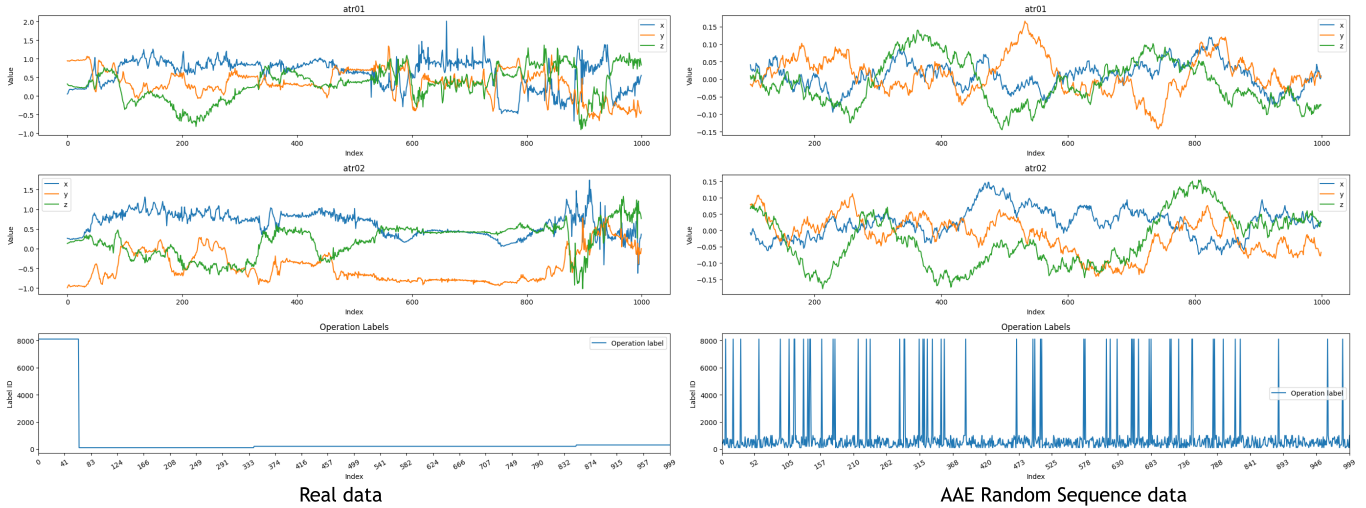}
    \caption{The figure presents the first 1000 data samples along with their corresponding operation labels. The left panel shows the raw data, while the right panel displays synthetic data generated using an AAE combined with a random sequence.}
    \label{fig:data_compare}
\end{figure*}

Figure \ref{fig:data_compare} illustrates a direct comparison between the original raw data (left panel) and the synthetic data (right panel) for the first 1000 samples, accompanied by their respective operation labels in the bottom plots. In the top and middle rows, the real signals (blue, orange, and green curves) exhibit varying levels of fluctuation and distinctive operational transitions, while the synthetic signals generated using an Adversarial Autoencoder (AAE) with a random sequence show comparable overall patterns but slightly different noise profiles and amplitude ranges. The operation labels, depicted in the bottom panels, provide a clear reference for identifying the corresponding operational states in both the raw and synthetic datasets, thereby enabling a visual assessment of how well the AAE-based generation approach captures the temporal behavior of the original system.
\section{Discussion}

\begin{figure*}[!ht]
    \centering
    \includegraphics[width=\linewidth]{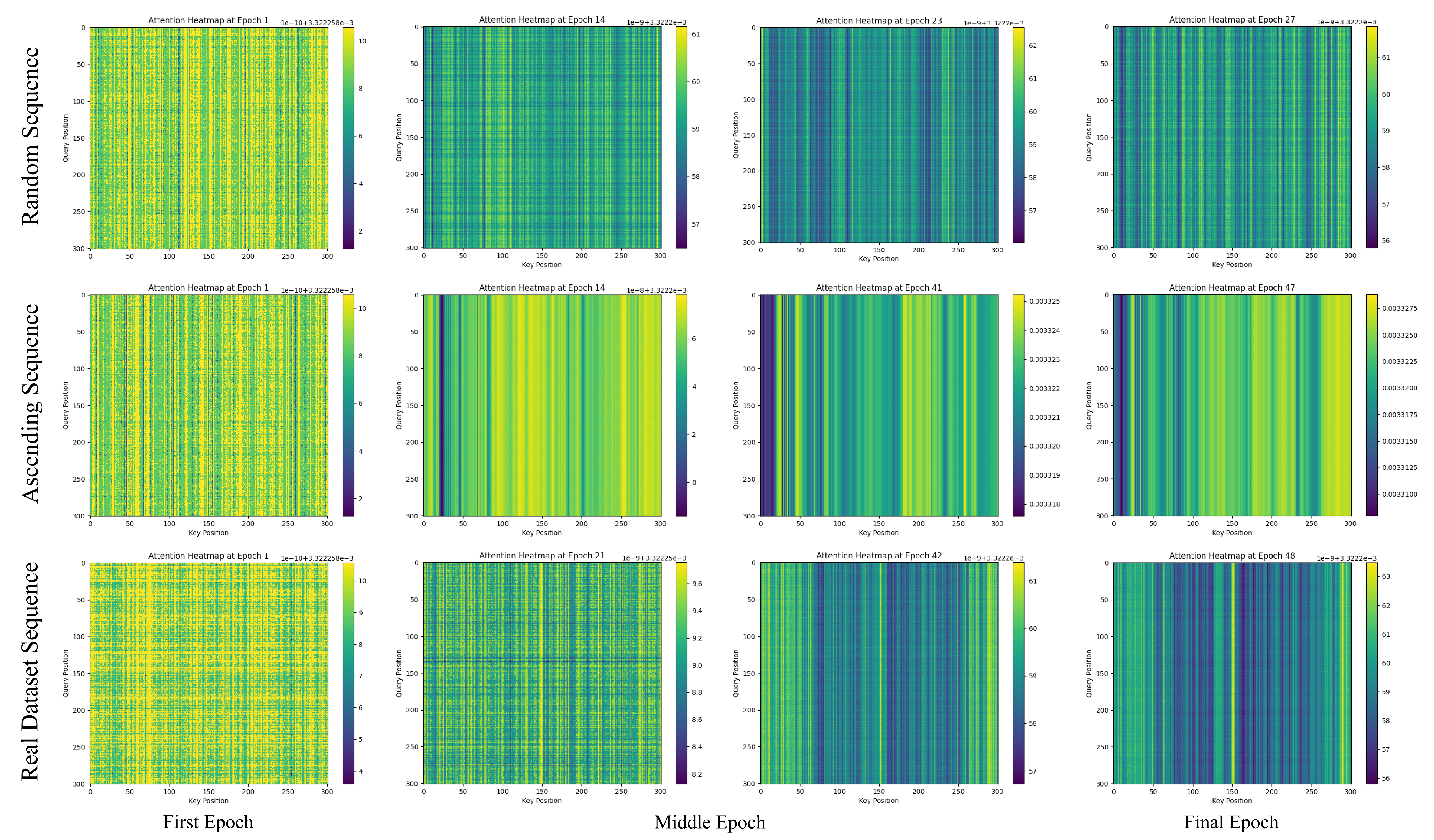}
    \caption{The figure of attention span in attention layers of discriminative model in 4 periods: initial, early mid, late mid, and end. The x-axis shows the key position value, and the y-axis shows the query position value. The figure shows that while the models work similarly initially, they start to get its characteristics through the training process. Specifically, the RS-augmented discriminative model tends to evaluate all the resources individually, while AS-augmented scanning for a long sequence and the RDSS one catch the feature by segments.}
    \label{fig:limit}
\end{figure*}

The findings of this study demonstrate the efficacy of a novel data augmentation strategy for industrial activity recognition, leveraging an Attention Autoencoder (AAE) in conjunction with random sequence reordering. Our findings show that a random order sequence can enhance the classification accuracy to 70\% and the F1 score to 64\%. Nonetheless, other augmentation data orders failed to promote the classification performance.

 Based on Table \ref{tab:my_table}, the random sequence outperformed other generated settings. Specifically, in the CTGAN synthetic set, the augmented data using this method improve the mean of classification accuracy by 2\%, whereas its counterparts failed to promote the result. Similarly, in AAE condition, RS leverage the F1 score of recognition task to 64\%. We discovered that in the random order case, the model has to switch the objective from sequence recognition to instant recognition (see Figure \ref{fig:limit}), which is caused by the nature of the random sequence. By performing this switch, the model gains immunity to the noise of activity changes in real-life scenarios.

 On the other hand, the ascending sequence failed to promote the classification performance. This outcome was the consequence of lacking transition in generated data. In other words, this setting makes the model scan for a clear, end-to-end sequence before making its decision, which is evident in figure \ref{fig:limit}. For example, the data sequence in real life can be the combination of three or more activities, which case is not generated at all in the AS setting. This confusion made by the lack of scenario in this setting leverage of more sophisticated settings for data augmentation in the future work, that it should cover a wide range of scenarios.

 It is also worth noting that the RDSS reconstruction method did not work as expected. In our discriminator attention layer analysis (see Figure \ref{fig:limit}), the RDSS setting sprays attention akwardly. In more detail, the RDSS attention span breaks into 3 notable parts - at the beginning, middle, and end of the sequence separately. We hypothesize that this segmentation arises because the data is frequently mixed across three sequences in this setting, prompting the model to focus on each section separately and vote on the outcomes accordingly.

\section{Conclusion}
\label{section:Conclusion}

This research showcases a random order approach in time series data augmentation, which was then compared with two other established methods. The study's contribution was built on an AAE and CTGAN to generate augmented data combined with a random permutation of the time-series sequences. The combination of random order and AAE demonstrated notable performance, achieving a statistically significant 64\% macro F1 score, while that of CTGAN yielded 70\% accuracy within the industrial activity recognition domain. These results open the door to new research pathways, such as investigating the underlying mechanisms that contribute to the success of random sequence reordering and exploring its applicability across diverse time-series datasets. Additionally, the research has shown that random reordering, when combined with an AAE, is the most characteristic for enhancing model performance in this context, while alternative methods like ascending sequence reconstruction and real data sequence reconstruction failed to yield comparable results. These findings emphasize the importance of this specific augmentation strategy and suggest that further research into its integration with generative models could lead to improved classification performance in time-series analysis.
\section* {Acknowledgment}
\label{Acknowledgment}

We acknowledge University of Technology (HCMUT), Vietnam National University Ho Chi Minh City (VNU-HCM) for supporting this study.

\bibliographystyle{ieeetr}

\begin{thebibliography}{10}

\bibitem{tee2022closelookhumanactivity}
W.~Z. Tee, R.~Dave, J.~Seliya, and M.~Vanamala, ``A close look into human activity recognition models using deep learning,'' in {\em 2022 3rd International Conference on Computing, Networks and Internet of Things (CNIOT)}, pp.~201--206, IEEE, 2022.

\bibitem{info10080245}
C.~Reining, F.~Niemann, F.~Moya~Rueda, G.~A. Fink, and M.~ten Hompel, ``Human activity recognition for production and logistics—a systematic literature review,'' {\em Information}, vol.~10, no.~8, 2019.

\bibitem{yoshimura_2023_8145223}
N.~Yoshimura, J.~Morales, and T.~Maekawa, ``Openpack: Public multi-modal dataset for packaging work recognition in logistics domain,'' July 2023.

\bibitem{10.1145/3675094.3678461}
H.~K. Phan, T.~N.~K. Nguyen, K.~C.~D. Nguyen, N.~P. Vo, A.~T. Ha, and N.~T. Le, ``Enhanced transportation and locomotion mode recognition through difference and variance analysis in inertial sensing data,'' in {\em Companion of the 2024 on ACM International Joint Conference on Pervasive and Ubiquitous Computing}, UbiComp '24, (New York, NY, USA), p.~585–590, Association for Computing Machinery, 2024.

\bibitem{8231736}
K.~Nurhanim, I.~Elamvazuthi, L.~I. Izhar, and T.~Ganesan, ``Classification of human activity based on smartphone inertial sensor using support vector machine,'' in {\em 2017 IEEE 3rd International Symposium in Robotics and Manufacturing Automation (ROMA)}, pp.~1--5, 2017.

\bibitem{8537550}
S.~Yu and L.~Qin, ``Human activity recognition with smartphone inertial sensors using bidir-lstm networks,'' in {\em 2018 3rd International Conference on Mechanical, Control and Computer Engineering (ICMCCE)}, pp.~219--224, 2018.

\bibitem{s21186316}
D.~Moreira, M.~Barandas, T.~Rocha, P.~Alves, R.~Santos, R.~Leonardo, P.~Vieira, and H.~Gamboa, ``Human activity recognition for indoor localization using smartphone inertial sensors,'' {\em Sensors}, vol.~21, no.~18, 2021.

\bibitem{8843403}
S.~W. Pienaar and R.~Malekian, ``Human activity recognition using lstm-rnn deep neural network architecture,'' in {\em 2019 IEEE 2nd Wireless Africa Conference (WAC)}, pp.~1--5, 2019.

\bibitem{leite2024transformerbasedapproachessensorbasedhuman}
C.~F. Souza~Leite, H.~Mauranen, A.~Zhanabatyrova, and Y.~Xiao, ``Transformer-based approaches for sensor-based human activity recognition: Opportunities and challenges,'' {\em Available at SSRN 5131703}, 2024.

\bibitem{8285168}
Z.~Wan, Y.~Zhang, and H.~He, ``Variational autoencoder based synthetic data generation for imbalanced learning,'' in {\em 2017 IEEE Symposium Series on Computational Intelligence (SSCI)}, pp.~1--7, 2017.

\bibitem{moshiri2020usingganenhanceaccuracy}
P.~F. Moshiri, H.~Navidan, R.~Shahbazian, S.~A. Ghorashi, and D.~Windridge, ``Using gan to enhance the accuracy of indoor human activity recognition,'' {\em arXiv preprint arXiv:2004.11228}, 2020.

\bibitem{xia2024preliminaryinvestigationsslcomplex}
Q.~Xia, T.~Maekawa, J.~Morales, T.~Hara, H.~Oshima, M.~Fukuda, and Y.~Namioka, ``Preliminary investigation of ssl for complex work activity recognition in industrial domain via moil,'' in {\em 2024 IEEE International Conference on Pervasive Computing and Communications Workshops and other Affiliated Events (PerCom Workshops)}, pp.~465--468, 2024.

\bibitem{10214902}
T.~Maekawa, Q.~Xia, R.~Otsuka, N.~Yoshimura, and K.~Tanigaki, ``Recent trends in sensor-based activity recognition,'' in {\em 2023 24th IEEE International Conference on Mobile Data Management (MDM)}, pp.~36--38, 2023.

\bibitem{xu2019modelingtabulardatausing}
L.~Xu, M.~Skoularidou, A.~Cuesta-Infante, and K.~Veeramachaneni, ``Modeling tabular data using conditional gan,'' {\em Advances in neural information processing systems}, vol.~32, 2019.

\bibitem{vaswani2023attentionneed}
A.~Vaswani, N.~Shazeer, N.~Parmar, J.~Uszkoreit, L.~Jones, A.~N. Gomez, {\L}.~Kaiser, and I.~Polosukhin, ``Attention is all you need,'' {\em Advances in neural information processing systems}, vol.~30, 2017.

\bibitem{SDV}
N.~Patki, R.~Wedge, and K.~Veeramachaneni, ``The synthetic data vault,'' in {\em IEEE International Conference on Data Science and Advanced Analytics (DSAA)}, pp.~399--410, Oct 2016.

\bibitem{michelucci2022introductionautoencoders}
U.~Michelucci, ``An introduction to autoencoders,'' {\em arXiv preprint arXiv:2201.03898}, 2022.

\end{thebibliography}

\end{document}